\documentclass[a4paper]{article}
\usepackage{epsfig}

\newcommand{\tensor}{\otimes}
\newcommand{\bra}[1]{\langle #1\vert}
\newcommand{\ket}[1]{|#1\rangle}

\newcommand{\braket}[2]{\langle #1\vert#2\rangle}

\newcommand{\unit}{\mathbf{1}}

\usepackage{amsfonts}
\usepackage{bbm}

\begin{document}


\title{Weak Values and Relational Generalisations}\author{Thomas Marlow\thanks{email: pmxtm@nottingham.ac.uk}\\ \emph{School of Mathematical Sciences, University of Nottingham,}\\
\emph{UK, NG7 2RD}}

\maketitle


\begin{abstract}
We justify generalisations of weak values from a tentatively relational perspective by deriving them from a generalisation of Bayes' rule.  We also argue that these generalisations have implications of quantum nonlocality and may form a novel approach to quantum gravity and cosmology.
\end{abstract}



\subsection*{Introductory Remarks}

Weak values are often invoked as a generalisation of expectation values in quantum theory \cite{AV90,AB05,Dios05}.  Curiously, however, such values can be outside the set of eigenvalues of their corresponding operators or even complex so, like Feynman's `negative probabilities' \cite{Feynman87}, such values cannot be interpreted strictly using orthodox statistics.  Nonetheless, they behave with a curious, and somehow consistent, logic \cite{ABPRT01}.  Without a clear interpretation or derivation it is hard to promote such weak values as fundamental or necessary; we need to derive that we \emph{must} use such things from a foundational standpoint.  Luckily, we have a convenient generalisation of weak values to hand: complex objects that we call `pedagogical examples' or `pegs' \cite{Marlow06c}.  This paper reviews what pegs are \cite{Marlow06c} and outlines why we use them, and why they are specifically a generalisation of weak values.  Pegs are axiomatically and relationally defined so we argue that they are natural in quantum theory---we do not use the orthodox ensemble interpretation of probability in order to derive them.  Thus we absolve any problems with frequentist derivations of weak probabilities that lie outside the interval $[0,1]$ or even outside the set of real numbers.  We will then discuss a few reasons why the use of orthodox statistics in quantum theory, and generalisations, is comparatively unnatural---using orthodox statistics we get quantum nonlocality, nonadditivity, and no clear path for tentative generalisations of quantum theory.

\subsection*{Weak Values and Pegs}

In the standard ensemble interpretation of quantum theory we pre-select an initial state and then work out the relative frequencies of results of measurements.  One can also retrodict by post-selecting information or one can both pre- and post-select information and infer about events or observables between such selections.  This is what weak values are; they are generalised expectation values of observables conditioned both on pre- and post-selected information.

In the orthodox frequentist interpretation of probabilities pre- and post-selection occurs by choosing to include only certain cases within a given ensemble.  One includes only those runs in which a particular initial state (a particular outcome of a previous measurement) was sent into the device and in which a particular outcome was received.  We can then try and make inferences about the values of observables between pre- and post-selection.  The weak value of an observable $A$ is defined to be

\begin{equation}
A_w = \frac{\bra{\Phi}\hat{A}\ket{\Psi}}{\braket{\Phi}{\Psi}}
\label{Weakvalue}
\end{equation}

\noindent where we have pre-selected a state $\ket{\Psi}$ and post-selected a state $\ket{\Phi}$.  Although these weak values can lie outside of the set of eigenvalues of the operator $\hat{A}$, they still seem to behave in a curiously consistent manner.  Clearly they cannot be interpreted using orthodox statistics---projection operators $\hat{P}$ are observables and thus can have weak values \emph{i.e.} one can have negative or complex weak probabilities.  Thus, if we are to derive or interpret weak values foundationally we must go outside of the orthodox statistical analysis because relative frequencies can always be placed in the interval $[0,1]$.  Note that the key idea in weak value theory is that weak values arise as a novel form of probabilistic inference \cite{ABPRT01}.  Luckily there is a form of statistics, distinct from orthodox statistics, that is framed in exactly this manner: Bayesian probability theory \cite{JaynesBOOK}.

Kastner has connected the ideas in histories formalisms to weak values \cite{Kastner03} although we wish to outline and derive this connection further.  So we wish to find a generalisation of weak probabilities in a histories formalism which are derived explicitly within the framework of Bayesian probabilistic inference.  We do not have to go far to find such a generalisation; we have postulated \cite{Marlow06b} and then derived \cite{Marlow06c} such pedagogical examples recently.  So, we can derive weak probabilities and generalisations from a Bayesian axiomatic standpoint without having to invoke dubious notions of negative or complex relative frequencies.  Let us summarise the basic results of \cite{Marlow06c} so as to explain what these pedagogical examples---`pegs'---are.

There exists a convenient histories formalism in the literature, namely the History Projection Operator (HPO) formalism introduced by Isham \cite{Isham94}.  One defines `history propositions' $\alpha, \beta, \gamma ... \in \cal{P(V)}$ which behave in ways analogous to projection operators (\emph{i.e.} there exist logical connectives analogous to the $\wedge, \vee$ and $\neg$ connectives in quantum logic)---$\cal{V}$ is a large Hilbert space and $\cal{P(V)}$ is its associated set of projection operators.  There is an additional natural logical connective that connects different temporal orderings which we call $\lhd$.  If we wish to promote a Bayesian approach then the simplest thing to do is simply to apply analogues of Cox's axioms of probability theory \cite{Cox46} to this histories propositional algebra.  Cox's approach is not the only formulation of Bayesian probability theory but it is the form we adopt here for pragmatic reasons (for a clear introduction to this form of Bayesian probability theory see \cite{JaynesBOOK}).  Cox originally gave two axioms of probability theory that are related to how the `and' and `negation' logical connectives behave in Boolean logic.  This is where we can bring in a relational spin to Cox's probability theory.  Relational theories are used to pragmatically ensure that factitious relationships are never introduced in physical theories.  In probability theory one does not wish to introduce any functional relationships between probabilities within the probability space that aren't rationally justified by the underlying propositional algebra.  This is exactly what Cox's axioms ensure.  Cox's $\wedge$-axiom is

\begin{equation}
p(\alpha \wedge \beta \vert I) := F[p(\alpha \vert \beta I), p(\beta \vert I)],
\label{COX1}
\end{equation}

\noindent where $F$ is an arbitrary function that is sufficiently well-behaved for our purposes.  Similarly, Cox's $\neg$-axiom is that the probability we assign to the negation of a proposition should only functionally depend upon the probability of the proposition before it was negated:

\begin{equation}
p(\neg \alpha \vert I) := G[p(\alpha \vert I)].
\label{COX2}
\end{equation}

These axioms ensure that we do not introduce any further functional relationships between the relevant probabilities (for we are not justified in doing so).  Of course we are being a bit na\"{\i}ve here because Cox applied these axioms to Boolean propositional algebras and not to quantum ones but, still, the key properties of the logical connectives are still satisfied:  the $\wedge$-connective is associative such that $\alpha \wedge (\beta \wedge \gamma) = (\alpha \wedge \beta) \wedge \gamma = \alpha \wedge \beta \wedge \gamma$ and the $\neg$-connective behaves such that if you apply it twice you get the same proposition back again.  We do not need an analogous $\vee$-axiom in the Boolean case because the logical connectives are logically dependent.  In quantum theories the `or' operation does not distribute with the `and' operation but, nonetheless, we can still na\"{\i}vely apply these axioms and see if we get something useful.

Note that when going from an algebra of propositions to an algebra of `history propositions' we naturally introduce a further logical connective, one that relates the different temporal orderings of the individual propositions within the `history proposition'.  We can define the operation $\lhd$ to reverse the temporal ordering (both dynamical and kinematical) of a history proposition.  This means that the $\lhd$-operation behaves like the $\neg$-operation in that if you apply it twice to a `history proposition' you get the same `history proposition' back again.  Thus we need an $\lhd$-axiom analogous to the $\neg$-axiom:

\begin{equation}
p(\lhd \alpha \vert I) := H[p(\alpha \vert I)],
\label{COX3}
\end{equation}

\noindent where $H$ is also an arbitrary function that is sufficiently well-behaved for our purposes.  One cannot always use real numbers for this Bayesian approach because the $\lhd$-axiom adds an `extra degree of freedom' to the associated probabilistic notions.  In the history of science we have all been rather vague about what we mean when we say the word `probability'.  It is a vague word that is often used in a variety of very specific ways which confuses the discussion.  Let us name these assignments $p(\alpha \vert I)$ `pedagogical examples' or `pegs' so as to be specifically vague rather than vaguely specific.  We have named them such for a variety of reasons.  Firstly we don't have to change the notation because `pegs' starts with a `p'.  Secondly, the name hints at a metaphor for the whole premise of the programme:  we are trying to the find the `right' peg to fit to corresponding hole.  We don't, for example, want to make the childish blunder of bashing a square peg into a round hole.  In \cite{Marlow06c} we argued that a corresponding set of pegs for the HPO algebra are complex numbers:

\begin{equation}
p(\alpha \vert I) = \mbox{tr}_{\cal{V}}(\alpha Y_I),
\label{Gleasonesque}
\end{equation} 

\noindent where the $Y_I$ operators are defined on the large Hilbert space ${\cal V}$.  We proved this using an analogue of Gleason's theorem which in turn was based on an analogue in the literature by Isham, Linden and Schreckenberg \cite{ILS94}.  This means that our complex pegs are in one-to-one correspondence with operators $Y_I$ on ${\cal V}$ (for dim ${\cal V} > 2$) in analogy to Gleason's theorem \cite{Gleason57} where probability assignments in the interval $[0,1]$ are in one-to-one correspondence with density matrices on ${\cal H}$ (for dim ${\cal H} > 2$).   For details see \cite{Marlow06c,ILS94}.  All we wish to emphasise here is that these pegs are derived axiomatically from a foundationally relational approach.  We do not need to use orthodox statistics to find them.  We now wish to show that these pegs are explicitly history generalisations of weak values of projection operators.

We define a homogeneous history proposition $\alpha$ as a time ordered tensor product of projection operators $\hat{\alpha}_{t_m} \in \cal{P(H)}$:

\begin{equation}
\alpha := \hat{\alpha}_{t_n} (t_n) \tensor \hat{\alpha}_{t_{n-1}}(t_{n-1}) \tensor ... \tensor \hat{\alpha}_{t_1}(t_1).
\end{equation}

\noindent  We stay in the Heisenberg picture such that each projection operator has the dynamics already implicit such that $\hat{\alpha}_{t_m} (t_m) = \hat{U}^\dagger(t_m - t_{m-1}) \hat{\alpha}_{t_m} \hat{U}(t_m - t_{m-1})$ where $\hat{\alpha}_{t_m}$ are Schr\"odinger picture projection operators.

Note that we can always rewrite tr$_{\cal H}(C_\alpha \hat{\rho})$ in the form of (\ref{Gleasonesque}) where the class operator $C_{\alpha}$  is given by $C_{\alpha} := \hat{\alpha}_{t_n}(t_n)\hat{\alpha}_{t_{n-1}}(t_{n-1})...\hat{\alpha}_{t_1}(t_1)$, ${\cal V} := \tensor^n {\cal H}$ and $\hat{\rho}$ is a density matrix on $\cal{H}$ \cite{Marlow06c,ILS94}.  So at least a subset of our pegs are of the form tr$_{\cal H}(C_\alpha \hat{\rho})$.  We can use this to connect our pegs to weak values.

If we apply (\ref{Weakvalue}) to projection operators we get

\begin{equation}
P_w = \frac{\bra{\Phi}\hat{P}\ket{\Psi}}{\braket{\Phi}{\Psi}}
\label{Weakprob}
\end{equation}

\noindent which could be termed a weak probability.  In general such weak probabilities are complex (also see \cite{Youssef94}) and they are related to our pegs in the following manner.  When we pre- and post-select then we are effectively making measurements and choosing to consider only a particular result.  We choose to represent such results with pure density matrices, or equivalently with projection operators.  Thus we can rewrite (\ref{Weakprob}) as a peg.  Let us take the history propostion $\alpha$ that consists of a series of three projection operators ordered in time.  We pre-select with $\hat{P}_{\ket{\Psi}} = \ket{\Psi}\bra{\Psi}$ and post-select with $\hat{P}_{\ket{\Phi}} = \ket{\Phi}\bra{\Phi}$.  Our peg for the history $\alpha = \hat{P}_{\ket{\Phi}} \tensor \hat{P} \tensor \hat{P}_{\ket{\Psi}}$ is

\begin{eqnarray}
p(\alpha \vert I) = \mbox{tr}_{\cal{H}}(C_\alpha) &=& \braket{\Psi}{\Phi} \bra{\Phi} \hat{P} \ket{\Psi} \\ &=& \braket{\Psi}{\Phi} P_w \braket{\Phi}{\Psi} \\ &=& \vert \braket{\Psi}{\Phi} \vert ^2  P_w.
\end{eqnarray}

\noindent Thus our pegs for three-time histories are explicitly trivially related to weak probabilities.  Pegs can be considered as generalisations of weak probabilities in a full histories formalism.

A compelling way to derive such weak values is through a generalisation of Bayes' rule \cite{Marlow06b}.  Note that 

\begin{eqnarray}
\vert \braket{\Psi}{\Phi} \vert ^2 = \mbox{tr}_{\cal H}(\hat{P}_{\ket{\Phi}}\hat{P}_{\ket{\Psi}}) = p(\alpha' \vert I) 
\end{eqnarray}

\noindent where $\alpha' = \hat{P}_{\ket{\Phi}} \tensor \hat{\unit} \tensor \hat{P}_{\ket{\Psi}}$.  Thus we have that

\begin{equation}
P_w = p(\alpha \vert \alpha' I) = \frac{p(\alpha \vert I)}{p(\alpha' \vert I)}
\end{equation}

\noindent such that $p(\alpha' \vert \alpha I) = 1$.  Hence weak probabilities can be explicitly derived by a conditionalisation using a generalisation of Bayes' rule!  Hence we have a generalised method to conditionalise pegs, although with novel results that need to be justified (this generalised Bayes' rule only behaves like one expects for commuting histories \cite{Marlow06b}).  This approach may allow us to conditionalise upon whole histories rather than just pre- and post-selection.

We have derived such entities through a relational Bayesian approach rather than use the problematic ensemble interpretation of probability.  There is no \emph{a priori} justification for weak values in an ensemble approach to probability.  Note that there are tentative physical justifications for weak values in terms of weak measurements \cite{AB05,Dios05}---we do not wish to contradict such an idea, rather we compliment it with a Bayesian generalisation of such objects so that we don't have to worry about nonsensical ideas like negative or complex relative frequencies.

Negative probabilities can be pragmatically justified \cite{Feynman87} by escaping the frequency approach but still they do not obey natural axioms of probability.  Note also that we do not contradict a relative frequency or decoherence approach through this work.  Such theories can be derived from ours by taking the real parts of pegs and enforcing consistency conditions so that they have the necessary limit stability properties \cite{Dios04}.  This is arguably what the consistent histories programme is about \cite{Griffith84,Omnes88,GH90}.  Similarly, the linearly positive histories formalism \cite{GP95} and Hartle's virtual probabilities formalism \cite{Hartle04} both use the real parts of pegs.

\subsection*{Quantum Theory Applications}

It is very interesting that this approach seems to be explicitly relational.  Leibniz's two principles of relationalism---the principle of sufficient reason and the principle of identifying the indiscernible---are very general and can apply as principles of physical science, not just to theories of spacetime.  One uses a relational approach simply for pragmatic reasons; it ensures that only rationally and physically justified elements are introduced in the theory---this helps in designing a theory which makes clear sense.  Thus relationalism is explicitly a philosophy about theory building and not an interpretation one takes for an existing theory.  We have argued that we can use this approach for generalising quantum theory by using analogues of Cox's axioms.  Cox's axioms explicitly ensure that we only introduce functional relations between probabilities when we are rationally justified in doing so (due to the underlying structure of the propositional algebra).

This may also have implications for quantum nonlocality.  We will present an analogy with general relativity.  General relativity is not a \emph{wholly} background independent theory \cite{Smolin05} but clearly Einstein used relationalism in order to derive it.  Einstein argued that the only elements in spacetime theories that are rationally justified are spacetime coincidences.  A relational approach must then ensure that one does not introduce relationships other than spacetime coincidences.  One wishes to `add the least' to the catalogue of spacetime coincidences \cite{Norton93}.  It is this reasoning that led him to general relativity.  We wish to do an analogous thing beginning with quantum theory and relationships between probabilities.  We can try and define probability theory relationally and thus naturally get to novel quantum probabilistic notions that are foundationally well-defined.  We wish to do such a thing because we do not consider the orthodox relative frequency approach suitable as we have problems in quantum theory because of it; namely nonadditivity \cite{Anast04}, nonlocality and infinite ensembles.

Relationalism helped Einstein bring locality back to gravitational physics \cite{Gisin05}, perhaps we can do an analogous thing with locality in quantum theory. Clearly we must tackle the issues that Bell brought to the table.

Bell proved that no hidden variable model (that obeys intuitive assumptions) can be used to \emph{emulate} the correlations we find in certain correlated quantum systems \cite{BellBOOK}.  If we postulate variables that pre-exist measurement that aren't effected by measurements or actions happening at spacelike separation then we can't, it seems, use such theories to emulate quantum theory.  This is why many refer to an implicit `nonlocality' in quantum theory.  To put it another way, he proved that quantum theory cannot be completed by using such hidden variables.  One can also validly argue that Bell proved that quantum theory itself cannot obey those same intuitive assumptions, so it either embodies causal nonlocalities or is not a \emph{hidden variable} completion of itself.  This is a result that goes against all our na\"{\i}ve assumptions about how the world works---a paradigm shifting body of work.

Of course, it is commonly accepted that Bell's analysis doesn't prove that quantum theory necessarily embodies \emph{causal} nonlocalities because it is not necessarily claimed that quantum theory is its own hidden variable completion.  It certainly isn't, as Bell amazingly proved, its own \emph{local} hidden variable completion (although Bell acknowledged that one might easily have qualms with his definition of locality \cite{BellBOOKsub}).  We can thus either accept nonlocality in quantum theory or we can try and approach quantum theory from a direction that necessarily ensures that we \emph{cannot} justifiably introduce hidden variables while making the same predictions; and then it tentatively \emph{might} be the case that we also get to keep locality.  Nonlocality might then be a factitious problem that is brought to the table \emph{by} bringing hidden variables to the table (while making the same predictions as quantum theory).  However, if one equates hidden variables with realism then the hidden variable hypothesis is hard to throw away.  Hopefully we will convince you that it is not so difficult to throw away at all.

From a relational perspective, a `complete' theory is one which explains all the physics by introducing only rationally justified theoretical relationships.  Thus perhaps we can still `complete' quantum theory.  This use of the word `complete' conflicts with the way that the word is used in terms of completing quantum mechanics by appending hidden variables.  Thus, to differentiate the two uses, we will write the relational use of the word `complete' in scare quotes.  We argue that there is, at present, sufficient ambiguity in the physics we do to justify either profound truth: nonlocality or `completeness'.

Let us first discuss how hidden variable theorists attempt to complete quantum theory and then we shall propose why and how we can `complete' it.  In quantum theory one cannot consistently assign values to observables prior to measurement.  This is arguably proved by the Bell-Kochen-Specker theorem \cite{BellBOOK,KS}.  We say `arguably' because recent valuations seem to nullify this theorem \cite{Meyer99,CK00}.  However, such valuations are necessarily so pathological so as to be experimentally unverifiable and thus we must be happy to remain ignorant of such valuations \cite{Apple04}.   Bayesianism is a way to remain happy in the light of this theorem because one explicitly \emph{presumes} that one is ignorant of such valuations---hence why we discuss probabilities.  In that sense quantum mechanics is trivially incomplete.  We wish to show more than this; quantum theory is necessarily incomplete because it is a theory of our ignorance.  If something is necessarily incomplete then why worry that it behaves in an odd manner when attempting to complete it?  Thus we argue that completing quantum theory, while maintaining its predictions, is a category error.  One must either derive a successor to quantum theory or `complete' it.  Instead of trying to complete a theory of ignorance by presuming more than we can rationally justify, we should `complete' our theory of ignorance by only presuming that that we can rationally justify while making all the relevant predictions.  We take a relational approach so as to attempt to `complete' quantum theory explicitly because relationalism is about designing physical theories such that one does not need to add anything to the catalogue of relations in order to make the relevant predictions---what else could we mean by the term `complete'?\footnote{A building is `complete' if it stands and functionally satisfies its design brief.  One could add any number of spurious features without effecting its function---flowers, art, monkey butlers and so forth---but such additions clearly do not `complete' the building even if they make it look pretty (this is not to say that art, flowers and monkey butlers can't be incorporated functionally into buildings, nor that our task shouldn't be to try and functionally incorporate such things; but if they were to be incorporated then we might manifestly have a new design brief and would probably wish to `complete' a different building with a different catalogue of funcitons).  Simply put, quantum theory stands and we can frame it such that it satisfies its minimal probabilistic catalogue of functions; it is a relational theory of ignorance.}  We need to make things as simple as possible---and not any simpler.

This is where the analogy with background independence might help; although note that it is clearly only an analogy.  We choose to use general relativity because it is partially background independent.  We choose to use background independent theories because we take a relational approach.  If we do choose a particular background then how are we to know whether the odd features (say nonlocality) of our theory are \emph{due} to that, seemingly arbitrary, choice?  We choose to identify these indiscernible backgrounds so that we don't end up invoking the vagaries of a particular arbitrarily chosen background.  Nor do we end up invoking the vagaries---if there are any---that come about by choosing \emph{any} particular background (vagaries that are universal to backgrounds).  If we can show that hidden variable theories are not rationally justified then we must, relationally, reject them.  (We could then argue that Cox's Bayesianism is to quantum mechanics as Leibniz's relationalism is to general relativity.  Both Cox's Bayesianism and Leibniz's relationalism consist of criteria of rationality.  Furthermore, since relationalism isn't necessarily a philosophy about space-time but rather a philosophy of science in general, we could say that Cox's theory is a relational theory of probability---although perhaps we confuse the issue by doing so.)  Clearly a particular background would complete general relativity.  However, it is a category error to complete general relativity in such a manner.  Hence we should rather call general relativity `complete' since we do not need to add anything to the catalogue of spacetime coincidences in order to make the relevant predictions.  This is not to say that no theory will ever superseed quantum theory or general relativity---we do not mean `complete' to be synonymous with `final'---we just use the word `complete' to emphasise that we should not complete theories in an irrational manner if they are rational (one should instead design a rational successor).  Perhaps there is a better word to use.

Now we need to present an argument that hidden variables are not rationally justified in a probability theory.  We will do so by introducing what we call demon observers, showing that invoking them is equivalent to invoking hidden variables, then emphasising that we have no rational reason to invoke them.

In quantum theory we cannot assign physical values to all variables, but we can assign probabilities to variables.  We explicitly assign probabilities to a variable \emph{because} we don't know which physical value, out of a given set of values, that variable takes.  If we did know which value the variable takes then we would not need a theory of probability.  Someone who \emph{does} know the value that a variable takes---particularly a value that a variable `hidden' from us takes---we call a demon observer.  Such demon observers live inside urns and measuring devices and they know the values of certain variables even if we don't.  We cannot rationally presume that these demon observers, as long as they are uncommunicative, effect our predictions, but it might be that they are a useful pedagogical device.  So, when we do an experiment we sometimes do not know what physical value a certain variable takes.  There are many possible values that it could take (in a counterfactual sense) and we can tentatively list them.

Often ignorance is associated with an ignorance as to which counterfactually distinct possibility is actually the case or equivalently an ignorance of what a demon observer sees.  However, knowing which value is the case alters the demon observer's probabilistic predictions (he clearly knows more than we do).  This is fine if we are ourselves the demon observer and come to know which value is the case; we just update our predictions.  However if we posit `hidden' variables then we come across a problem---these demon observers are uncommunicative.  There is no reason to presume that a theory of a demon observer's ignorance is operationally equivalent to a theory of our ignorance---demon observes know far more than we do (the values of hidden variables or their prior-probabilities for example).  Demon observers who know the hidden configuration will predict different probabilities for future (or past) events in comparison to an observer who does not know the `hidden' configuration.  Thus we must distinguish theories that involve demon observers from those that involve a single no-nonsense observer.  It is clear that the former is irrational (we invoke little imp-like observers who we cannot communicate with) and the later obeys simple plausible criteria of rationality.  Bell proved that if we can communicate with these demon observers, while making the same predictions, then we must get some form of nonlocality---however, it is utterly plausible, one might say inevitable, that if we were able to communicate with demon observers then we would manifestly make different predictions.  Hence we nullify any curious claims that Bell's analysis logically implies reality, or quantum theory, necessarily embodies causal nonlocalities.  By far the more controversial claim (locality being plausible but tentative), is the claim that a probability theory can be `complete'.

Thus it seems that we reject realism but this is not the case.  Probability theory is a theory of ignorance so we explicitly assume that we cannot assign values to things (hence we are profoundly happy with the B-K-S theorem).  In probability theories we explicitly assume that variables have values---we just don't know what they are and we cannot consistently assign values to things while making the same probabilistic predictions.  If we were to assume that variables \emph{don't} have values then we would not even have the basis for a probability theory! We might be able to assign values to things by making novel predictions, so some theory might yet superseed quantum theory to shine light on such hidden variables---we have nothing against hidden variable theories in this sense.  It is tautological however that quantum theory is not its own successor and it is vacuous to call quantum theory incomplete because of this tautology---every theory would then be incomplete.  It does not necessarily follow that quantum theory can be `completed' but we argue that it can because it can be framed as a rational theory of our ignorance using pegs (\emph{cf.} our analogy with background independence).  While writing this paper we found another that makes similar strong claims \cite{SR06} (and it is interesting that such claims are also made within a relational approach).

Having recognised the domain of Bell's work, many authors have attempted to extend and generalise the discussion without having to use the probabilistic assumptions that Bell used.  Some good examples are the exciting programmes started by Hardy \cite{Hardy92} and Stapp \cite{Stapp03}.  Stapp's programme, for example, is in opposition to the tentative view given above; it is an attempt to prove that causal nonlocalities are necessarily embodied by quantum theory.  Stapp's work, however, has recently been criticised by Shimony \cite{Shimon04}.  One could argue that probabilities are not well-defined independently of events in space-like separated regions---they are manifestly defined with respect to such regions.  (Note the distinction between a probability that is well-defined due to an ignorance of spacelike separated events and a probability that is independent of knowledge of spacelike separated events; the latter is ill-defined.) Similarly, Shimony argues that the counterfactual statements that Stapp invokes are also not well-defined independently of space-like separated regions---counterfactual statements are manifestly defined relative to such regions.  Such interdependencies between the very definitions of certain probabilities or between the very definitions of certain counterfactual statements may be interpreted in an entirely logical manner---thus we need not \emph{necessarily} invoke nonlocal causal relations to explain them.  Thus it is not yet clear that such programmes prove that causal nonlocalities are necessarily a component of quantum theory.  Of course, all this relates to the perennial debate started by the conflict between the EPR paper \cite{EPR} and Bohr's response \cite{Bohr35}.

Clearly we haven't argued that quantum theory is a local theory.  We are just not convinced that it necessarily embodies causal nonlocalities.  Bell's profound analysis tells us that, if we wish to tentatively deny nonlocality, we need to either approach quantum theory from a novel perspective or make novel predictions; we have attempted to do the former and leave the latter for a profound revolution in sub-quantum physics.  For further relational ideas in regards to quantum theory see \cite{Rovel96,Poulin05,Jaros05}.  Perhaps it seems that we have been very dismissive of hidden variable theories; note rather that we have been very conservative in arguing only that hidden variables should not be used in deciding whether a probability theory is `complete' (by definition a probability theory is one in which we \emph{do not} assign realistic values to those variables that are `hidden' from us---hence we assign probabilities instead).

\subsection*{Quantum Gravity Applications}

Our Bayesian approach is also itself ripe for generalisation.  One takes a propositional algebra and derives pegs relationally using the logical connectives of the algebra.  With a novel propositional algebra, say in quantum gravity or quantum cosmology, one will derive novel pegs.  Thus we have a tentatively constructive approach.

Above we presented an analogy between Leibniz's relationalism and Cox's Bayesianism.  However we think that there is something deeper than just an analogy here.  Note for example that hidden variable theories require a particular reference frame/background \cite{Hardy92}.  It might be that approaching quantum theory in a way analogous to relationalism we might make some first tentative steps on an approach to quantum gravity.  It is commonly considered a pre-requisite that theories of quantum gravity are more relational than general relativity (we try and remove as much background structure as possible \cite{Smolin05}).  Note that this is not to say that quantum gravity must uniquely be relational.  Once we have found quantum gravity it might be that we could interpret it in a variety of ways \cite{Rickles05}.  Relationalism is an approach we take for opportunistic reasons, it is explicitly a pragmatic philosophy of theory building and we physicists are theory builders; we remain silent about how we are to interpret the theory in the end for that is something we cannot speak of yet.  We thus consider Cox's Bayesian probability theory (or rather relational generalisations of probability theory) to also be a pre-requisite.  In the light of quantum cosmology we cannot use orthodox statistics anyway because we cannot make an ensemble of our universe to test except as a pedagogical device---Bayesianism is, at present, the only real alternative to orthodox statistics and Bayesianism is, in the least, honest about using pedagogical devices.

In order to tentatively begin to apply these relational probabilistic ideas to quantum gravity we need a propositional algebra on which to apply analogues of Cox's axioms.  There are a few possibilities in the literature.  Markopoulou's acausal sets \cite{Marko00} and Savvidou's covariant histories \cite{Savvid05} are two prime examples.  We need a set of propositions which involve logical connectives.  If we have analogues of the four we used above we will perhaps end up using complex numbers, or discrete analogues of complex numbers.  But we may have more connectives in general, perhaps involving an explicit notion of causality.  There also may exist generalisations of Bayesian probability theory that apply to Heyting algebras which might also be applicable in cosmological reasoning \cite{Marko99}---this we intend to investigate.  These ideas are so tentative we do not wish to comment on them further except to note that Hardy has recently introduced a framework in which one can have both relational spacetime and probabilistic concepts from the outset \cite{Hardy05}.  It is perhaps not \emph{a priori} valid to assume that quantum gravity will be a probabilistic theory, however this is our working hypothesis.

One might even be able to take the analogy with relationalism further.  We have previously introduced complex assignments but perhaps we should only discuss the rational relationships between assignments.  Any assignments that obey Cox's axioms would do to an extent, but by choosing a particular type of assignment one might introduce vagaries particular to those assignments, and we should identify them all because we have no reason to choose one type of assignment over another (they are indiscernibles).  This might help with issues we have in regards to the \emph{a priori} use of real and complex numbers in quantum theory \cite{Isham02}.  We hope to investigate this tentative analogy between Leibniz's relationalism and Cox's Bayesianism further in future work.

There are teasing links between this approach and toposophic approaches in the literature \cite{IB98,IB99} based around a reformulation of the Bell-Kochen-Specker theorem.  Instead of attempting (and failing) to assign values to variables in quantum theory one can assign generalised valuations in an attempt to maintain some form of realism in quantum theory.  This is particularly desired in quantum cosmology.  These generalised valuations maintain functional relationships between observables and the logical relationships between propositions.  In \cite{IB98} the logical structure is different to what we have assumed above and there is a further element of contextuality.  However, the analogy with relationalism may still hold for generalised valuations.  Pegs and probabilities are generalised valuations of sorts and it might be useful to use them in such an approach.

\subsection*{Conclusions}

We conclude that weak values and generalisations---which we call pegs---can be derived from a Coxian analysis.  This tentatively absolves problems we have with deriving such notions, and with deriving probabilities themselves, using orthodox statistics.  We argue that we can give a relational interpretation of such pegs;  this provides an analogy between Leibniz's relationalism in spacetime and Cox's approach to probability theory that we can use to tentatively deny quantum nonlocality.  We argue that this approach is ripe for generalisation in that one could begin to apply these ideas to novel propositional algebras---it is not clear how to apply orthodox quantum theory to such things.  Perhaps we could apply a pragmatic relational notion of probability to a relational, but propositional, notion of spacetime.  In a certain limit (or limits), we might then be able to get standard quantum theory and general relativity back again (for very tentative discussions about such limits see \cite{MS04} and \cite{Catich05} respectively).

\subsection*{Acknowledgements}

We would like to thank George Jaroszkiewicz for the original prompt to investigate further the relationship between pegs and weak values, and for all his help throughout my studies.  Thanks are also due to EPSRC for funding this work.


\begin{thebibliography}{99}

\bibitem{AV90} Y. Aharonov and L. Vaidman, ``Properties of a quantum system during the time interval between two measurements'' \emph{Phys. Rev. A} \textbf{41} (1990) 11--20.

\bibitem{AB05} Y. Aharonov and A. Botero, ``Quantum Averages of Weak Values'' \emph{Phys. Rev. A} \textbf{72} (2005) 052111, preprint: {\tt quant-ph/0503227 v2}.

\bibitem{Dios05} L. Di\'{o}si, ``Weak Measurements in Quantum Mechanics'' (2005) preprint: {\tt quant-ph/0505075 v1}.

\bibitem{Feynman87} R. P. Feynman, ``Negative Probabilities'' in \emph{Quantum Implications}, Eds., B. J. Hiley and F. D. Peat (Routledge and Kegan Paul, 1987) 235--248.

\bibitem{ABPRT01} Y. Aharonov, A. Botero, S. Popescu, B. Reznik and J. Tollaksen, ``Revisiting Hardy's Paradox: Counterfactual Statements, Real Measurements, Entanglement and Weak Values'' \emph{Phys. Lett. A} \textbf{301} (2002) 130--138, preprint: {\tt quant-ph/0104062 v1}.

\bibitem{Marlow06c} T. Marlow, ``A Bayesian Analogue of Gleason's Theorem'' (2006) preprint: {\tt quant-ph/0603065}.

\bibitem{JaynesBOOK} E. T. Jaynes, \emph{Probability Theory: The Logic of Science} (Cambridge University Press, 2003).

\bibitem{Kastner03} R. E. Kastner, ``Weak Values and Consistent Histories in Quantum Theory'' \emph{Stud. Hist. Phil.
Mod. Phys.} \textbf{35} (2004) 57--71, preprint: {\tt quant-ph/0207182 v3}.

\bibitem{Marlow06b}  T. Marlow, ``Bayesian Probabilities and the Histories Algebra'' to be published by \emph{Int. Jour. Theo. Phys.} (Accepted, 2006) preprint: {\tt gr-qc/0603011}.

\bibitem{Isham94} C. J. Isham, ``Quantum logic and the history propositions approach to quantum theory'' \emph{Jour. Math. Phys.} \textbf{35} (1994) 2157--2185, preprint: {\tt gr-qc/9308006}.

\bibitem{CoxBOOK} R. T. Cox, \emph{The Algebra of Probable Inference} (The Johns Hopkins University Press, 1961).

\bibitem{Cox46} R. T. Cox, ``Probability, frequency, and reasonable expectation'' \emph{American Jour. Phys.} \textbf{14} (1946) 1--13.

\bibitem{ILS94}  C. J. Isham, N. Linden and S. Schreckenberg ``The Classification of Decoherence Functionals: An Analogue of Gleason's Theorem'' \emph{Jour. Math. Phys.} \textbf{35} (1994) 6360--6370, preprint: {\tt gr-qc/9406015}.

\bibitem{Gleason57} A. M. Gleason, ``Measures on the Closed Subspaces of a Hilbert Space'' \emph{Jour. Math and Mech.} \textbf{6} (1957) 885--893.

\bibitem{Youssef94} S. Youssef, ``Quantum Mechanics as Complex Probability Theory'' \emph{Mod. Phys. Lett A} \textbf{9} (1994) 2571.

\bibitem{Dios04} L. Di\'{o}si, ``Anomalies of weakened decoherence criteria for quantum histories'' \emph{Phys. Rev. Lett.} \textbf{92} (2004) 170401, preprint: {\tt quant-ph/0310181 v1}.

\bibitem{Griffith84} R. B. Griffiths, ``Consistent history propositions and the Interpretation of Quantum Mechanics'' \emph{Jour. Stat. Phys.} \textbf{36} (1984) 219--273.

\bibitem{Omnes88} R. Omn\'es, ``Logical reformulation of quantum mechanics. I. Foundations'' \emph{Jour. Stat. Phys.} \textbf{53} (1988) 933--955.

\bibitem{GH90} M. Gell-Mann and J. B. Hartle, ``Quantum Mechanics in the light of quantum cosmology'' in \emph{Proceedings of the Third International Symposium on the Foundations of Quantum
Mechanics in the Light of New Technology} (Physical Society of Japan, Tokyo, Japan, 1990) 321--343.

\bibitem{GP95} S. Goldstein and D. N. Page ``Linearly Positive history propositions: Probabilities for a Robust Family of Sequences of Quantum Events'' \emph{Phys. Rev. Lett.} \textbf{74} (1995) 3715--3719,  preprint: {\tt gr-qc/9403055}.

\bibitem{Hartle04} J. B. Hartle, ``Linear Positivity and Virtual Probability'' \emph{Phys. Rev. A} \textbf{70} (2004) 02210, preprint: {\tt quant-ph/0401108}.

\bibitem{Smolin05}  L. Smolin, ``The case for background independence'' (2005) preprint: {\tt hep-th/0507235}.

\bibitem{Norton93} J. D. Norton, ``General covariance and the foundations of general relativity: eight decades of dispute'' \emph{Rep. Prog. Phys.} \textbf{56} (1993) 791--858.

\bibitem{Anast04} C. Anastopoulos, ``On the relation between quantum mechanical probabilities and event frequencies'' 
\emph{Annals. Phys.} \textbf{313} (2004) 368, preprint: {\tt quant-ph/0403207}.

\bibitem{Gisin05} N. Gisin, ``Can relativity be considered complete?  From Newtonian nonlocality to quantum nonlocality and beyond'' (2005) preprint: {\tt quant-ph/0512168 v1}.

\bibitem{BellBOOK} J. S. Bell, \emph{Speakable and unspeakable in quantum mechanics}, (Cambridge University Press, 2004).

\bibitem{BellBOOKsub} J. S. Bell, ``The theory of local beables'' in \cite{BellBOOK}.

\bibitem{KS} S. Kochen and E. P. Specker, ``The problem of hidden variables in quantum mechanics'' \emph{Jour. Math. Mech.} \textbf{17} (1967) 59--87.

\bibitem{Meyer99} D. A. Meyer, ``Finite precision measurement nullifies the Kochen-Specker theorem'' \emph{Phys. Rev. Lett.} \textbf{83} (1999) 3751--3754, preprint: {\tt quant-ph/9905080}.

\bibitem{CK00} R. Clifton and A. Kent, ``Simulating Quantum Mechanics by Non-Contextual Hidden Variables'' \emph{Proc. Roy. Soc. Lond. A} \textbf{456} (2000) 2101--2114, preprint: {\tt quant-ph/9908031}.

\bibitem{Apple04} D. M. Appleby, ``The Bell-Kochen-Specker Theorem'' \emph{Stud. Hist. Phil. Mod. Phys.} \textbf{36} (2005) 1, preprint: {\tt quant-ph/0308114 v1}.

\bibitem{SR06} M. Smerlak and C. Rovelli, ``Relational EPR'' (2006) preprint: {\tt quant-ph/0604064 v1}.

\bibitem{Hardy92} L. Hardy, ``Quantum Mechanics, local realistic theories, and Lorentz invariant realistic theories''  \emph{Phys. Rev. Lett.} \textbf{68} (1992) 2981--2984.

\bibitem{Stapp03} H. P. Stapp, ``A Bell-type theorem without hidden variables'' \emph{American Jour. Phys.} \textbf{72} (2004) 30--33.

\bibitem{Shimon04} A. Shimony, ``An Analysis of Stapp's ``A Bell-type theorem without hidden variables'''' (2004) preprint: {\tt quant-ph/0404121}.

\bibitem{EPR} A. Einstein, B. Podolsky and N. Rosen, ``Can Quantum-Mechanical Description of Physical Reality Be Considered Complete?''  \emph{Phys. Rev.} \textbf{47} (1935) 777.

\bibitem{Bohr35} N. Bohr, ``Can Quantum Mechanical Description of Physical Reality be Considered Complete?'' \emph{Phys. Rev.} \textbf{48} (1935) 696--702.

\bibitem{Rovel96} C. Rovelli, ``Relational Quantum Mechanics'' \emph{Int. Jour. Theo. Phys.} \textbf{35} (1996) 1637--1678, preprint:  {\tt quant-ph/9609002}.

\bibitem{Poulin05} D. Poulin, ``Toy Model for a Relational Formulation of Quantum Theory'' (2005) preprint:  {\tt quant-ph/0505081 v2}.

\bibitem{Jaros05} G. Jaroszkiewicz, ``Quantized Detector Networks: a Quantum Informational Approach to the Description and Interpretation of Quantum Physics'' (2005) preprint: {\tt quant-ph/0511196}.

\bibitem{Rickles05} D. Rickles, ``A new spin on the hole argument''  \emph{Stud. Hist. Phil. Mod. Phys.} \textbf{36}  (2005) 691.

\bibitem{Marko00} F. Markopoulou, ``Quantum Causal Histories'' \emph{Class. Quant. Grav.} \textbf{17} (2000) 2059--2072, preprint: {\tt hep-th/9904009 v5}.

\bibitem{Savvid05} N. Savvidou, ``General Relativity Histories Theory'' \emph{Braz. Jour. Phys.} \textbf{35} (2005) 307--315, preprint: {\tt gr-qc/0412059}.

\bibitem{Marko99} F. Markopoulou, ``The internal description of a causal set: What the universe looks like from th inside'' \emph{Commun. Math. Phys.} \textbf{211} (2000) 559--583, preprint: {\tt gr-qc/9811053 v2}.

\bibitem{Hardy05} L. Hardy, ``Probability Theories with Dynamic Causal Structure:  A New Framework for Quantum Gravity'' (2005) preprint: {\tt gr-qc/0509120 v1}.

\bibitem{Isham02} C. J. Isham, ``Some Reflections on the Status of Conventional Quantum Theory when Applied to Quantum Gravity'' (2002) preprint: {\tt quant-ph/0206090 v1}.

\bibitem{IB98} C. J. Isham and J. Butterfield, ``A Topos perspective on the Kochen-Specker Theorem: I. Quantum States as Generalised Valuations'' (1998) preprint: {\tt quant-ph/9803055 v4}.

\bibitem{IB99} C. J. Isham and J. Butterfield, ``Some Possible Roles for Topos Theory in Quantum Theory and Quantum Gravity'' \emph{Found. Phys.} \textbf{30} (2000) 1707-1735, preprint: {\tt gr-qc/9910005 v1}.

\bibitem{MS04} F. Markopoulou and L. Smolin, ``Quantum Theory from Quantum Gravity'' (2004) preprint: {\tt gr-qc/0311059 v2}.

\bibitem{Catich05} A. Caticha, ``The Information Geometry of Space and Time'' (2005) preprint:  {\tt gr-qc/0508108 v1}.


All preprints refer to the http://arxiv.org/ website.

\end{thebibliography}
\end{document}